\documentclass[conference,a4paper]{IEEEtran}
\IEEEoverridecommandlockouts
\usepackage{array}
\usepackage{booktabs}
\usepackage{cite}
\usepackage[font=small]{caption}
\usepackage{amsmath,amssymb,amsfonts}
\usepackage{algorithmic}
\usepackage{graphicx}
\usepackage{textcomp}
\usepackage{xcolor}
\usepackage{subcaption}
\usepackage{todonotes}
\usepackage{microtype}
\def\BibTeX{{\rm B\kern-.05em{\sc i\kern-.025em b}\kern-.08em
    T\kern-.1667em\lower.7ex\hbox{E}\kern-.125emX}}

\usepackage{multirow}
\usepackage {url}
\usepackage [numbers,sort&compress]{natbib}
\usepackage[super]{nth}
\usepackage{paralist}
\usepackage[a4paper, total={7in, 9.53in}]{geometry}
\usepackage{color}

\newcolumntype{L}[1]{>{\raggedright\arraybackslash}m{#1}}
\newcolumntype{C}[1]{>{\centering\arraybackslash}m{#1}}

\makeatletter

\def\ps@IEEEtitlepagestyle{%
  \def\@oddfoot{\mycopyrightnotice}%
  \def\@evenfoot{}%
}

\newcommand\copyrighttext{%
  \footnotesize \textcopyright~2019 IEEE. Personal use of this material is permitted. Permission from IEEE must be obtained for all other uses, in any current or future media, including reprinting/republishing this material for advertising or promotional purposes, creating new collective works, for resale or redistribution to servers or lists, or reuse of any copyrighted component of this work in other works.}

\def\mycopyrightnotice{%
  {\fbox{\parbox{\dimexpr\textwidth-\fboxsep-\fboxrule\relax}{\copyrighttext}}}
  \gdef\mycopyrightnotice{}
}

\begin{document}

\title{On the Cost-Optimality Trade-off for Service Function Chain Reconfiguration}

\author{
\IEEEauthorblockN{Kyoomars Alizadeh Noghani, Andreas Kassler, and Javid Taheri\\
\IEEEauthorblockA{\small Karlstad University, Karlstad, Sweden\\ Email: \{kyoomars.noghani-alizadeh, andreas.kassler, javid.taheri\}@kau.se}}}


\maketitle


\begin{abstract}
Optimal placement of Virtual Network Functions (VNFs) in virtualized data centers enhances the overall performance of Service Function Chains (SFCs) and decreases the operational costs for mobile network operators. Maintaining an optimal placement of VNFs under changing load requires a dynamic reconfiguration that includes adding or removing VNF instances, changing the resource allocation of VNFs, and re-routing corresponding service flows. However, such reconfiguration may lead to notable service disruptions and impose additional overhead on the VNF infrastructure, especially when reconfiguration entails state or VNF migration. On the other hand, not changing the existing placement may lead to high operational costs. In this paper, we investigate the trade-off between the reconfiguration of SFCs and the optimality of the resulting placement and service flow (re)routing. We model different reconfiguration costs related to the migration of stateful VNFs and solve a joint optimization problem that aims to minimize both the total cost of the VNF placement and the reconfiguration cost necessary for repairing a suboptimal placement. Numerical results show that a small number of reconfiguration operations can significantly reduce the operational cost of the VNF infrastructure; however, too much reconfiguration may not pay off should heavy costs be involved.
\end{abstract}


\section{Introduction}
\label{sec:intro}
Network Function Virtualization (NFV) has been an important driver for the deployment of next-generation networks such as 5G in modern virtualized data centers. In NFV, expensive proprietary network boxes such as firewalls or video transcoders, which are deployed on customized hardware, are virtualized and shipped as Virtual Machines (VMs), containers or microservices. Such Virtual Network Functions (VNFs) can be placed on commodity servers utilizing the flexibility and scalability aspects of modern virtualized data centers. Forming a set of Service Function Chains (SFCs), multiple VNFs cooperate by exchanging traffic to provide a given end-to-end service.

A major challenge is where (on which servers) to place such VNFs and how to route the traffic in the substrate network, because the placement determines the performance, operational cost, and resiliency of the SFCs. In this regard, there has been a considerable effort in optimizing such placement and many optimization models and heuristics have been proposed in the literature~\cite{7444915}. However, all placement decisions that might be optimal, at a given point in time, may become suboptimal as consequent placement decisions embed new SFCs into the substrate network without changing the placement of existing VNFs. Consequently, it is required to dynamically reconfigure the SFC placement and service flow routing over time to improve the optimality of the placement. Additionally, when objectives change, a reconfiguration might be needed as different placements of VNFs in the network realize different objectives (e.g., energy optimal versus resilient). Finally, reconfiguration of the placement of SFC helps network providers to achieve other objectives, for example, reducing resource contentions, resolving constraint violations, and conducting a set of scheduled actions such as server or switch maintenance in a data center.

Reconfiguration is, however, a challenging task for network providers as it may lead to instability, service disruptions, and performance degradation. This problem is exacerbated for SFCs as a service disruption caused by the reconfiguration of a VNF may lead to severe performance degradation of following nodes, and consequently the chain as a whole. Moreover, reconfiguration imposes overhead to the network and servers. For example, relocation of stateful VNFs encompasses the migration of system states of a VNF from one server to another, which consumes network resources and imposes additional CPU stress to both source and destination end-hosts. 

Reconfiguration of SFCs entails relocating, provisioning, scaling, terminating, and merging instances of VNF(s) as well as re-routing their corresponding service flows. The overheads of SFC reconfiguration are considered in several works. For instance, authors in~\cite{tajiki} considered the difference in the routing matrix before and after the reconfiguration as the cost. Authors in~\cite{7335318} characterized the migration cost by a penalty value. Carpio~\textit{et al.}~\cite{8406275} considered the service disruption due to VNF migration as a cost. Tang~\textit{et al.}~\cite{8338124} modeled the reconfiguration cost by the time that it takes to migrate a VNF instance and and Eramo~\textit{et al.}~\cite{7866881} represented the reconfiguration cost by the revenue loss during the time that the VNF is unable to provide a service due to migration. Although considering the effort required to change the placement of VNFs is crucial, a single parameter for the reconfiguration cost does not represent the complexity and overhead of reconfiguration. In addition, parameters such as duration of migration as well as the migration downtime vary based on the traffic load of the network, and thus should be modeled more accurately rather than assuming a fixed value as in~\cite{8338124, 7866881}. Furthermore, the amount of efforts (cost) required to perform the reconfiguration operations may hamper the gain of an optimal VNF placement. As a result, it is important to investigate the trade-off between reconfiguration cost and optimality of the resulting VNF placement after the reconfiguration has been performed. Such trade-off between (optimality vs. reconfiguration costs) in the context of SFCs is not thoroughly investigated yet.

In this paper, we formulated several reconfiguration costs for SFCs when using stateful VNFs. The reconfiguration costs are added as an extension to an existing optimal VNF placement model to form a joint optimization problem modeled as an Integer NonLinear Program (INLP). Using this model, we study the trade-off between the gain achieved by a better target placement of VNFs and the reconfiguration cost required to realize such placement, given an existing placement of VNFs. Numerical evaluations show that a minimum number of reconfiguration operations can significantly improve the optimality of the placement; however, making the placement completely optimal in terms of a certain objective may require a prohibitive amount of reconfiguration operations. Additionally, results show that considering the reconfiguration cost in finding an optimal placement may lead to better reconfiguration strategies with lower overheads.
 
The rest of the paper is organized as follows. Section~\ref{sec:system-model} defines the objective function, identifies input parameters, introduces decision variables, and formulates reconfiguration costs of SFCs. Selected numerical results are presented in Section~\ref{sec:eval}. The paper is concluded in Section~\ref{sec:conclusion}.




\section{VNF Reconfiguration}
\label{sec:system-model}
\subsection{Objective Function}
Given a current SFC placement and service flow routing, the problem is to jointly minimize a generic objective and the total reconfiguration cost as follows:
\begin{align}
\label{eq:general-objective-func}
\text{(P):}&~~{\text{minimize:}} (1 - \alpha)\underbrace{(Cost_{NP})}_{(1)} + ~(\alpha)\underbrace{(Cost_{REC})}_{(2)}.
\end{align}

The optimization problem (P) is composed of two normalized components: (1) the cost of the new placement, and (2) the reconfiguration cost. The former is the cost for the new SFC placement and service flow routing (e.g., in terms of total energy consumption of the VNF infrastructure) \emph{after reconfiguration} and the latter is the cost required to achieve the new placement from the current one, where potentially some VNFs have been migrated and some flows have been rerouted. By weighting the two cost components with the parameter $0 \le \alpha\le 1$, we can study different trade-offs. A small $\alpha$ tries to obtain new placements and flow routing that aim to optimize the NFV infrastructure regarding a given cost function; it may lead to many VNF migrations and flow rerouting. A large $\alpha$, on the other hand, prioritizes target placements that can be achieved with a low reconfiguration cost. 

In the optimization problem (P), we focus in detail on the following aspects: (1) modeling reconfiguration costs (part $(2)$ of (P)), and (2) investigating the trade-off between reconfiguration costs and optimality. As we do not focus on part $(1)$ of (P), we adopt an existing model from~\cite{tajiki} that has the objective to reduce the energy consumption of the VNF infrastructure. Please note, that part $(1)$ of the optimization problem (P) can be replaced by any other cost functions proposed in the literature.

\subsection{Input Parameters and Decision Variables}
We consider a network with $N$ switches, $X$ servers, $S$ SFCs, $V$ different types of VNFs, and $F$ service flows. We denote the current routing matrix for placed service flows in the network with $M_{N\times N}^{F}$ matrix as follows:
\begin{equation}
   M_{i,j}^{f}=\left\{
   \begin{array}{ll}
   1, & \hbox{if flow $f$ traverses the link} \\
         & \hbox{between switch $i$ and switch $j$,} \\
   0, & \hbox{otherwise.}
   \end{array}
   \right.
\end{equation}

When a VNF is migrated from the source server $X_1$ to the destination server $X_2$, we assume that the VNF migration traffic (used to transfer CPU, memory, etc.) is routed on the shortest path. It is possible that a given link belongs to multiple shortest paths. We also assume that a portion of the bandwidth, denoted by $BW$, is reserved for VNF migration traffic on all the links on the shortest path. This information is required as they are used later to determine the duration of migration because the migrating VNFs need to share the reserved capacity for migration traffic. The matrix $K$ indicates which pair of servers share a common link on their shortest path and is defined as:
\begin{equation}
   K_{x,y}^{z,w}=\left\{
   \begin{array}{ll}
   1, & \hbox{if the shortest path between} \\
         & \hbox{servers ($x$,$y$) and servers ($z$,$w$)} \\
         & \hbox{share a common link,} \\
   0, & \hbox{otherwise.}
   \end{array}
   \right.
\end{equation}

Size of flow $f$ and VNF of type $v$ are denoted by $\lambda_{f}$ and $A_{v}$ vectors. A VNF of type $v$ dedicates $L_{v}$ of its processing resource to process a unit of a flow. Utilization cost and maximum processing capacity of server $x$ are denoted by $\psi_{x}$ and $C_{x}$, respectively. Finally, the matrix $U_{x}^{v,s}$ identifies the current placement of VNFs (for each SFC) on the servers. For instance, $U_{2}^{1,1} = 1$ expresses that SFC $1$ has a VNF of type $1$, which is currently placed at the server $2$. Likewise, $U_{x,f}^{v,s}$ identifies where a flow that belongs to a SFC receives service from a VNF.

The model takes the following inputs: (1) a given placement of VNFs on the servers, and (2) their corresponding flow routing. Then, the model calculates a new placement and flow routing (as given by the decision variables in Table~\ref{table:decision-variables}) to minimize the joint operational cost of the new placement and the reconfiguration cost required to achieve it.

\begin{table}[htb]
  \caption{List of decision variables.}
  \centering
  \label{table:decision-variables}
      \begin{footnotesize}
    \begin{tabular}{ |>{\centering\arraybackslash} p{1cm} |>{\centering\arraybackslash} p{0.6cm} |>{\centering\arraybackslash} p{1cm} | p{4.1cm} |}
    \hline
    Decision Var. &  Type & Size & Description \\ \hline
    $R_{i,j}^{f}$ & \{0,1\} & $F.N^{2}$ & New routing matrix; `1' means flow $f$ crosses link $(i, j)$ \\ \hline
    $W_{x}^{v,s}$ & \{0,1\} & $X.V.S$ & New VNF placement; `1' if VNF $v$ of SFC $s$ is allocated at server $x$ \\ \hline
    $W_{x,f}^{v,s}$ & \{0,1\} & $F.X.V.S$ & New flow allocation; `1' if flow $f$ uses VNF $v$ of SFC $s$ at server $x$ \\ \hline
    \end{tabular}
    \end{footnotesize}
    \vspace{-4mm}
\end{table}

\subsection{Reconfiguration Costs}
\textbf{Number of Flow Rerouting}: When the current and the new placement of a VNF are different, its corresponding traffic must be rerouted. This requires the installation of new forwarding rules in the network and the termination of obsoleted ones. Increasing the number of rerouted flows may result in network instability, the increase of packet loss, degraded throughput, and the increase of end-to-end delay~\cite{tajiki, Gandhi:2017}. Eq.~\eqref{eq:rerouted-flows} determines the number of routing rules that need to be changed (as a result of the network reconfiguration).
\begin{equation}
\label{eq:rerouted-flows}
\mathcal{U} = \sum_{i=1}^{N} \sum_{j=1}^{N} \sum_{f=1}^{F} |R_{i,j}^{f}-M_{i,j}^{f}|.
\end{equation}

\textbf{Migration Size}: Migrating a stateful VNF imposes overhead to the underlying network because it generates extra network traffic to transfer states between the source and the destination hosts during the migration. Therefore, from the networking point of view, migration of a VNF with fewer states to transfer is more preferred. 

In this paper, we assume that each VNF is deployed over a separate VM. As a result, relocation of a stateful VNF becomes similar to VM migration using well-known live VM migration schemes including pre-copy~\cite{clark2005live}, post-copy~\cite{Hines:2009}, and hybrid~\cite{sahni2012hybrid} when only CPU and memory states migrate between the source and destination and the disc is shared. The total migration traffic size varies based on the migration scheme used. In hybrid migration, the upper bound on migration traffic is two times the memory plus the disc size. In pre-copy migration, the page dirty rate plays a major role~\cite{remedy-vijay}. However, in post-copy migration the traffic comprises mainly the disc size and the size of the memory. Although post-copy is inherently less robust against failures, it can provide predictable behavior and possibility to estimate the longest possible migration time. Therefore, we assume that the post-copy migration scheme is used. The overhead of reconfiguration due to migration of stateful VNF is presented in Eq.~\eqref{eq:VNF-size}.
\begin{equation}
\label{eq:VNF-size}
\mathcal{V} = \sum_{x=1}^{X} \sum_{s=1}^{S} \sum_{v=1}^{V} W_{x}^{v,s}(1 - U_{x}^{v,s})A_{v}.
\end{equation}

The decision variable $W_{x}^{v,s}$ in conjunction with the input parameter $U_{x}^{v,s}$ determines if the new placement of the VNF is different from its current placement.

\textbf{Migration Time}: In addition to the size of the migration traffic, total migration time is an important factor in evaluating the performance of migration. Ideally, the total migration time ought to be short to prevent performance degradation and minimize the overhead of migration on the underlying network and end-hosts. For instance, the post-copy migration scheme first transmits all processor state to the target, starts the VM at the target, and then actively pushes the VM's memory pages from source to the target. The effectiveness of the post-copy scheme depends on its ability to minimize the number of page-faults by pushing the pages from the source before they are faulted by the VM at the target~\cite{Hines:2009}. However, if the duration of migration is prolonged (e.g., due to congestion in the network), the probability of page faults increases and the performance of the migrated VNF degrades once it resumes service on the target host.

\begin{figure}[t]
\begin{centering}
\includegraphics[width=0.45\textwidth]{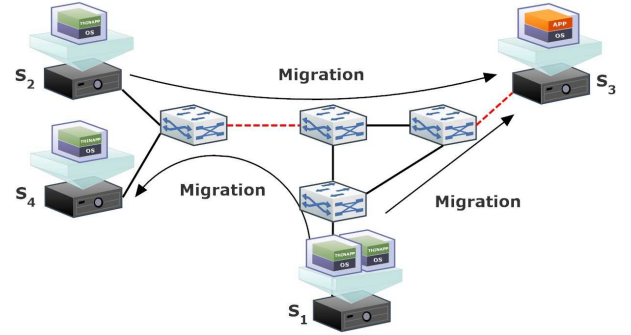}
\par\end{centering}
\caption{\label{fig:bottlenecklink}Possible bottleneck links due to simultaneous migrations.}
\vspace{-5mm}
\end{figure}

One of the important parameters that determines the duration of the migration is the amount of bandwidth reserved for the migration traffic~\cite{MigrationCost}. The bottleneck link between two servers with the minimum amount of bandwidth (i.e., carrying the maximum number of parallel VNF migration flows) determines the total migration duration for a VNF together with the volume of migration traffic. The main challenge is to find the bottleneck link. The bottleneck between servers $x$ and $y$ can be (1) a link between the source and destination server and their corresponding first-hop switches and/or (2) any of the links on the shortest path between the source and destination servers. Let us denote these two types of links by $\eta_{x}$ and $\eta_{x,y}$, respectively. The source and destination links may become the bottleneck when shortest paths between a pair of servers do not share a common link. In contrast, a link on the shortest path may become a bottleneck when it is shared amongst multiple shortest paths used for migration. Fig.~\ref{fig:bottlenecklink} depicts the bottleneck links (shown with dashed lines), for three simultaneous VNF migrations between servers \textless$S_{1}$,$S_{4}$\textgreater, \textless$S_{1}$,$S_{3}$\textgreater, and \textless$S_{2}$,$S_{3}$\textgreater.

Eq.~\eqref{eq:source-destination-link} calculates the number of migrations to/from a given server and Eq.~\eqref{eq:middle-link} determines the number of migrations among each pair of servers. The total number of migrations carried on a link between servers $(x,y)$ is equal to the total number of migrations between servers $(x,y)$ plus the total number of migrations between all pairs of servers that have a common link with shortest paths between servers $(x,y)$.
\begin{subequations}
\begin{align}
& \eta_{x} = \sum_{s = 1}^{S} \sum_{v = 1}^{V} |W_{x}^{v,s} - U_{x}^{v,s}|,\qquad\qquad\qquad\qquad~~\forall x,\label{eq:source-destination-link}\\
\begin{split}
& \eta_{x,y} = \sum_{s = 1}^{S} \sum_{v = 1}^{V} W_{x}^{v,s} (1 - U_{y}^{v,s}) +\\
& \qquad{}\sum_{z = 1}^{X} \sum_{w = 1}^{X} \sum_{s = 1}^{S} \sum_{v = 1}^{V} K_{x,y}^{z,w} W_{z}^{v,s} (1 - U_{w}^{v,s}),\quad~\forall x,y.\label{eq:middle-link} 
\end{split}
\end{align}
\end{subequations}

We assume that the bandwidth set aside for the migration traffic is shared equally among all migration traffics passing through the bottleneck link. In order to make the model tractable, we do not reclaim the bandwidth for the migration traffic once a VNF is migrated. As a result, the time for live migration of a given VNF, using the post-copy migration scheme, can be estimated using Eq.~\eqref{eq:migration-duration-one-VNF}.
\begin{equation}
\label{eq:migration-duration-one-VNF}
T_{x,y}^{v,s} = \frac {W_{x}^{v,s}(U_{y}^{v,s})} {\frac {BW}{~\underset{(x,y) \in X}{\max}(\eta_{x}, \eta_{x,y}, \eta_{y})}},
\end{equation}
where the $max$ function identifies the link with the highest number of ongoing migrations. Finally, as we assume that all migrations start in parallel, the total migration time is equal to the longest migration time as calculated in Eq.~\eqref{eq:migration-duration}.
\begin{equation}
\label{eq:migration-duration}
\mathcal{W} = ~\underset{(s,v,x,y)}{\max}(T_{x,y}^{v,s} A_{v}),
\end{equation}
where $v \in V$, $s \in S$, and $x,y \in X$.

Note that the total migration time in other migration schemes, in particular pre-copy scheme, depends not only on the bandwidth allocated for the migration traffic but also on additional factors including the page dirty rate and the migration completion deadline.

\textbf{Migration Downtime}: The duration of downtime is another key parameter that determines the quality of any live migration. The downtime is the time period that the migrating VNF does not respond, because either its critical states are in transition from the source to the destination end-host or the network is not yet converged~\cite{7776571, 8459946}. For network operators, the duration of downtime is equal to revenue loss and must be minimized. Eq.~\eqref{eq:revenue-loss} formulates the revenue loss for a post-copy migration procedure and all SFCs.
\begin{equation}
\label{eq:revenue-loss}
\mathcal{X} = \sum_{s=1}^{S} ~\underset{(x,y) \in X}{\max}(W_{x}^{v,s}U_{y}^{v,s}) P_{s}F_{s} \rho, \qquad\qquad\qquad\forall v,
\end{equation}
where $P_{s}$ characterizes the revenue loss of the SFC $s$ per one Gbit of lost traffic, $F_{s}$ denotes the total size of flows belong to the SFC $s$, and $\rho$ is a small constant to represent the time required to transfer a minimal subset of a VNF's execution states (e.g., CPU states and registers). Finally, $~\underset{(x,y) \in X}{\max}(W_{x}^{v,s}U_{y}^{v,s})$ determines if any of VNFs in SFC $s$ is migrated. 

\textbf{VNF Utilization}: The memory size and memory access pattern of a VNF before the migration is one of the key parameters that determines the duration of service disruption that includes the migration duration and downtime. Because VNFs typically perform memory or I/O intensive operations, the intensity of their loads increases with the number of their flows. As in~\cite{8406275}, we penalize the migration of a VNF in proportion to its utilization before the migration. The rationale behind this idea is to minimize the probability and impact of service disruption. Even if the duration of service disruption is the same for two distinct VNFs, it is preferred to migrate the VNF that is serving fewer flows (assuming that all flows have the same priority). Eq.~\eqref{eq:vnf-util-sub} calculates the utilization of a VNF before the migration. 
\begin{equation}
\label{eq:vnf-util-sub}
u_{s,v} = \sum_{f=1}^{F} \frac  {{U_{x,f}^{v,s}} \lambda_{f} {L_{v}}} {C_{x}},\qquad\qquad\qquad\qquad\quad~\forall v, s, x.
\end{equation}

Additionally, the network operators may desire to facilitate or prevent migrating specific VNFs regardless of their utilization. This goal can be achieved by assigning a separate penalty value to each VNF type as expressed in Eq.~\eqref{eq:vnf-util-sub-penalty}.
\begin{equation}
\label{eq:vnf-util-sub-penalty}
QoSPenalty_{s,v} = W_{x}^{v,s} (1 - U_{x}^{v,s}) \kappa_{v},\qquad\quad\forall v, s, x,
\end{equation}
where $\kappa_{v}$ is a parameter that can be tuned to determine the penalty of a migration for VNF of type $v$, and $W_{x}^{v,s}(1-U_{x}^{v,s})$ determines the migration. Finally, Eq.~\eqref{eq:vnf-util} calculates the total cost of QoS degradation for all SFCs.
\begin{equation}
\label{eq:vnf-util}
\mathcal{Y} = \sum_{s=1}^{S} \sum_{v = 1}^{V} u_{s,v}QoSPenalty_{s,v}.
\end{equation}

\textbf{Server Overhead}: Conducting a migration imposes additional stress on the source and destination end-hosts as they have to dedicate a portion of their resources to the migration process (e.g., CPU, memory, and network)~\cite{isci2011improving}. Eq.~\eqref{eq:server-overhead} considers the overhead of migration for all servers.
\begin{equation}
\label{eq:server-overhead}
\mathcal{Z} = \sum_{x=1}^{X} \sum_{s=1}^{S} \sum_{v = 1}^{V} |W_{x}^{v,s} - U_{x}^{v,s}|\psi_{x}.
\end{equation}

The value of the overhead cost for each server, denoted by $\psi_{x}$, is the key parameter to determine the reconfiguration cost associated with it. For instance, consider a scenario where the administrator needs to exclude a server from a reconfiguration procedure to avoid imposing additional overhead to the specific part of the network that a given server is located. This goal can be easily achieved by assigning a higher overhead value to that server.

\textbf{Total Reconfiguration Cost:} $Cost_{REC}$ denotes the total cost for SFC reconfiguration as the summation of all the aforementioned costs; that is:
\begin{equation}
\label{eq:total-cost}
Cost_{REC} = \mathcal{U} + \mathcal{V} + \mathcal{W} + \mathcal{X} + \mathcal{Y} + \mathcal{Z}.
\end{equation}

All reconfiguration costs in~\eqref{eq:total-cost} are normalized. In addition, we linearized  Eq.~\eqref{eq:migration-duration-one-VNF}, Eq.~\eqref{eq:migration-duration}, and Eq.~\eqref{eq:revenue-loss}. We omit the details of the linearization process of the formulated INLP (Integer Non-Linear Programming) due to lack of space. Once the problem is linearized, the joint optimization problem (P) can be optimally solved using an off-the-shelf software package. The problem to solve is NP-hard. However, as we assume that we are dealing with a planning approach that does not require real-time solutions, complexity does not impose a significant concern.

\section{Numerical Results}
\label{sec:eval}
We conduct numerical evaluations to explore the trade-off between reconfiguration cost and optimality. The linearized problem is solved using the MATLAB toolbox YALMIP~\cite{1393890} and the GUROBI~\cite{gurobi} solver.


\begin{figure*}[t]
    \centering
    \begin{subfigure}[b]{0.47\textwidth}
        \centering
        \includegraphics[width=\textwidth]{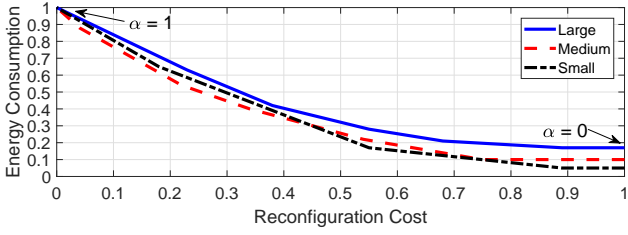}
        \caption[]%
        {{\small Normalized energy consumption versus reconfiguration cost.}}  
        \label{fig:trade-off-normalized}
    \end{subfigure}
    \hfill
    \begin{subfigure}[b]{0.47\textwidth}
        \centering
        \includegraphics[width=\textwidth]{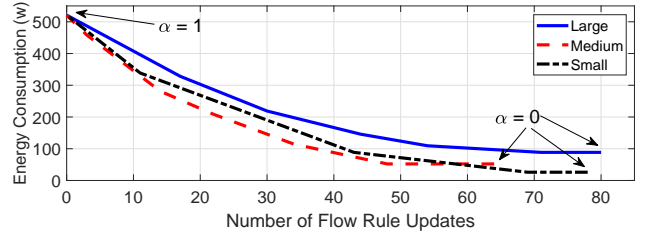}
        \caption[]%
        {\small Energy consumption versus number of updated flow rules.}
        \label{fig:trade-off-flow}
    \end{subfigure}
    \caption[]
    {Trade-off between energy consumption and reconfiguration cost for the network with $15$ switches and $10$ servers.}
    \label{fig:tradeoff}
    \vspace{-3mm}
\end{figure*}

Three scenarios with different numbers of SFCs, VNFs, and flows are considered. We assume that servers have equal processing and storage capacities. However, the energy consumption and overhead cost of servers are different. SFCs have the same number of VNFs and penalty costs. VNFs are different in size, but similar in CPU requirements and penalty values. Moreover, flows are different in sizes and QoS requirements and are uniformly distributed between SFCs. The main characteristics of all scenarios, SFCs, VNFs, servers, and flows for the following experiments are presented in Tables~\ref{table:scenarios} and~\ref{table:input-characteristics} where numbers in brackets define a range.

All scenarios are deployed over a leaf-spine network topology with $5$ spine switches, $10$ leaf switches, and $10$ servers. All links have 10 Gbps bandwidth and 1ms latency. The scenarios as well as the topology are selected for tractability purposes due to the computational complexity of the problem (P).
\begin{table}[t]
  \vspace{-1mm}
  \caption{Characteristics of different scenarios.}
  \centering
  \label{table:scenarios}
      \begin{footnotesize}
    \begin{tabular}{ |>{\centering\arraybackslash} p{1.5cm} |>{\centering\arraybackslash} p{1.5cm} |>{\centering\arraybackslash} p{2cm} |>{\centering\arraybackslash} p{1.5cm} |}
    \hline
    &  \# SFCs & \# VNFs per SFC & \# Flows \\ \hline
    Large & 15 & 4 & 20 \\ \hline
    Medium & 10 & 3 & 15 \\ \hline
    Small & 5 & 3 & 10 \\ \hline
    \end{tabular}
    \end{footnotesize}
\end{table}
\begin{table}[t]
  \caption{Input parameters for SFCs, VNFs, servers, and flows.}
  \centering
  \label{table:input-characteristics}
    \begin{footnotesize}
    \begin{tabular}{ |>{\centering\arraybackslash} p{0.75cm} |>{\centering\arraybackslash} p{4.5cm} |>{\centering\arraybackslash} p{1.75cm}|}
\hline
                        & Parameter                     & Value \\ \hline
\multirow{1}{*}{SFC}    & Revenue loss per Gbit per second (\$)  & 500 \\ \hline
\multirow{4}{*}{VNF}    & Num. CPU required  & 1 \\ \cline{2-3} 
                        & Memory required (GB)  & {[}1-2{]} \\ \cline{2-3} 
                        & CPU load ratio per unit of flow (hz)   & 100 \\ \cline{2-3} 
                        & No service penalty   & 1 \\ \hline
\multirow{5}{*}{Server} & CPU (Ghz)            & 2  \\ \cline{2-3} 
                        & Memory (GB)          & 50 \\ \cline{2-3} 
                        & Num. CPU cores       & 16 \\ \cline{2-3} 
                        & Power usage (W)       & {[}20-90{]} \\ \cline{2-3} 
                        & Overhead penalty     & {[}20-50{]} \\ \hline
\multirow{2}{*}{Flow}   & Size (MB)            & {[}50-100{]} \\ \cline{2-3} 
                        & Delay threshold (ms) & {[}50-100{]} \\ \hline
    \end{tabular}
    \end{footnotesize}
    \vspace{-4mm}
\end{table}

VNFs are initially placed on servers such that the placement is non-optimal, however, constraints are not violated. For instance, if the objective (1) in the optimization problem (P) is to minimize the energy, we initially distribute VNFs over all servers to maximize the energy consumption, e.g., all servers are powered ON. Such non-optimal placement may emerge in reality when VNFs are decommissioned or new ones are deployed without changing the placement of existing ones.

To investigate the trade-off between reconfiguration cost and optimality of the SFC placement and routing after the reconfiguration has been applied, we vary the value of $\alpha$. Fig.~\ref{fig:tradeoff} contains two demonstrations of the trade-off between the optimality of the new placement and flow routing and reconfiguration cost to achieve the target placement when $\alpha$ is decreased from $1$ by $0.1$ in each step. Fig.~\ref{fig:trade-off-normalized} displays the normalized energy consumption of the target placement compared with the normalized reconfiguration cost. The main reason for such a demonstration is that reconfiguration is composed of different costs with different units. Fig.~\ref{fig:trade-off-flow}, on the other hand, represents the trade-off between energy consumption and a specific reconfiguration overhead (flow rule updates) to provide the reader with insight about the actual values. Similar figures could have been plotted for energy consumption and  other reconfiguration costs explained in Section~\ref{sec:system-model}. However, they are not presented due to space restrictions.

Recall from the optimization problem (P), $\alpha = 0$ gives the most energy-efficient solution and $\alpha = 1$ leads to the most reconfiguration efficient target placement. As can be seen in Fig.~\ref{fig:trade-off-normalized}, for a large value of $\alpha$, reconfiguration efforts are close to optimal ($0$ on the x-axis), while the new placement is being less energy optimal. In contrast, the reconfiguration cost is high for a small $\alpha$ since the target placement tends to not consider the reconfiguration cost required to achieve the target placement. Instead, it enforces the most energy optimal placement, leading to a potentially high number of reconfiguration operations. Considering Fig.~\ref{fig:tradeoff}, we observes the following outcomes.

\begin{figure}[t]
    \begin{centering}
    \includegraphics[width=0.45\textwidth]{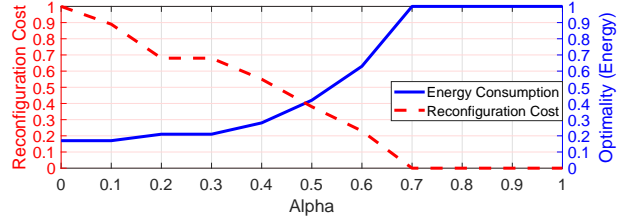}
    \par\end{centering}
    \caption{Trade-off detail for the large scenario.}
    \label{fig:tradeoff-detail}
    \vspace{-4mm}
\end{figure}

\textbf{Considering reconfiguration cost is important}: The same level of optimality can be achieved with different amount of reconfiguration effort. This conclusion can be seen in Fig.~\ref{fig:trade-off-normalized} where, for example, in large scenario the level of energy consumption is constant while the value of reconfiguration cost increases from $0.9$ to $1$. Another representation of the results for the large scenario is depicted in Fig.~\ref{fig:tradeoff-detail}. As it can be seen, the energy consumption $0.17$ is achieved with two different amounts of reconfiguration effort: $1$ for $\alpha = 0$ and $0.89$ for $\alpha = 0.1$. This outlines the fact that considering reconfiguration cost in finding an optimal placement is crucial. While for $\alpha = 0$ the optimization problem finds the target placement by blindly selecting VNFs to move and changing their corresponding routing matrix, the optimization problem reaches the same level of optimality at $\alpha = 0.1$ by using a better reconfiguration strategy that leads to less reconfiguration overhead.

\begin{figure*}[t]
    \centering
    \begin{subfigure}[b]{0.47\textwidth}
        \centering
        \includegraphics[width=\textwidth]{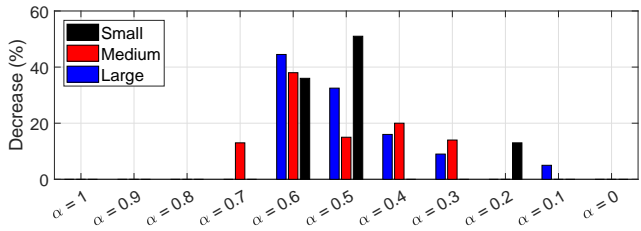}
        \caption[]%
        {{\small Decreased energy consumption.}}  
        \label{fig:energy-topo-2}
    \end{subfigure}
    \hfill
    \begin{subfigure}[b]{0.47\textwidth}
        \centering
        \includegraphics[width=\textwidth]{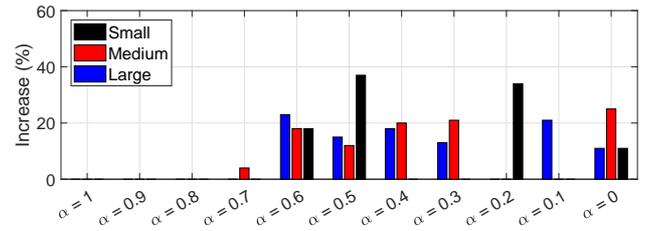}
        \caption[]%
        {\small Increased reconfiguration cost.}
        \label{fig:reconf-topo-2}
    \end{subfigure}
    \caption[]
    {Percentage of reduced energy consumption and increased reconfiguration cost by varying the value of $\alpha$ for all scenarios.}
    \label{fig:improvement}
    \vspace{-4mm}
\end{figure*}

\textbf{Reconfiguration may not pay off}: Another important observation is that by accepting a small reconfiguration cost, the initial placement can be significantly improved in terms of energy cost. Adding more reconfigurations (using smaller $\alpha$) may not necessarily reduce the energy cost. For all scenarios, Fig.~\ref{fig:improvement} shows the percentage of the reduced energy consumption (Fig.~\ref{fig:energy-topo-2}) and the increased reconfiguration cost (Fig.~\ref{fig:reconf-topo-2}) at each step in proportion to its previous step. As expected, for $\alpha = 1$ the energy consumption is not reduced as the optimization problem only considers the overhead of reconfiguration. By shifting the focus from reconfiguration efficient solution to a more energy-efficient one (reducing alpha from $1$ to $0$) the amount of energy consumption starts to decrease. However, significant improvements are achieved when reconfiguration and optimality have the same importance in the optimization problem. As it can be seen in Fig.~\ref{fig:improvement}, for $\alpha = 0.5$, the energy consumption is reduced by $73\%$ on average (in comparison to energy consumption level at $\alpha = 1$) with the average $42\%$ increase in the reconfiguration cost. Further attempts to find a better energy-efficient solution leads to smaller improvements at the expense of relatively high reconfiguration cost. For instance, at $\alpha = 0$ the energy consumption for all scenarios, is reduced by the remaining $27\%$ on average at the cost of $58\%$ increase in the reconfiguration cost.

\section{Conclusion and Future Work}
\label{sec:conclusion}
In this paper, we investigated the trade-off between the cost of optimality and reconfiguration in the context of stateful VNF placement. Different reconfiguration costs are considered and formulated to include the size of the VNF to migrate, time of migration, overhead on servers, QoS degradation as well as the revenue loss due to information loss occurring during the migrations. Our numerical evaluation showed that with a small number of reconfiguration steps, a suboptimal placement can be repaired leading to a significant cost reduction for the energy to run the NFV infrastructure. In contrast, making a suboptimal placement optimal may entail a large number of reconfigurations, which may not pay off at the end. Furthermore, numerical evaluations showed that the same level of optimality can be achieved with different amount of reconfiguration costs. This issue highlights the fact that considering the overhead of reconfiguration in finding an optimal placement is crucial, as it may help to design a better reconfiguration strategy. 

However, these conclusions are not comprehensive as the reconfiguration cost is significantly influenced by the initial placement of VNFs, the target state, as well as the network and SFC topology. Therefore, even a small change in any of the aforementioned parameters can drastically change the reconfiguration cost. Although this paper sheds some lights on the trade-off between optimality and the amount of effort or cost that network operators have to accept to realize such placement, it is just the first step to evaluate such trade-offs. In the future extension of this work, we intend to investigate this trade-off over a wider range of scenarios including different network typologies, VNF settings, and SFC workloads. We will also develop fast multi-objective heuristics to solve the stated  optimization problem.

\section*{Acknowledgment}

This work has been funded by the Knowledge Foundation Sweden through the profile HITS.

{\footnotesize{}\bibliographystyle{./IEEEtran}
\bibliography{refs}
}{\small \par}

\end{document}